\documentstyle[aaspp]{article}
\tighten
\begin{document}
\def\etal{{\it et al.\/}}
\def\cf{{\it cf.\/}}
\def\ie{{\it i.e.\/}}
\def\eg{{\it e.g.\/}}

\title{A new instability of accretion disks around magnetic compact stars}
\author{{\bf Mario Vietri\altaffilmark{1} and Luigi Stella\altaffilmark{2,3}}}
\altaffiltext{1}{Universit\`a di Roma 3, Via della Vasca Navale 84, 
00147 Roma, Italy, e-mail: vietri@corelli.fis.uniroma3.it }
\altaffiltext{2}{Osservatorio Astronomico di Roma, Via dell'Osservatorio 2, 
00040 Monte Porzio Catone (Roma), Italy, 
e-mail: stella@coma.mporzio.astro.it} 
\altaffiltext{3}{Affiliated to the International Center for Relativistic
Astrophysics}

\begin{abstract}
Aperiodic variability and Quasi Periodic Oscillations (QPOs) are observed 
from accretion disks orbiting white dwarfs, neutron stars, and black holes, 
suggesting that the flow is universally broken up into discrete blobs.
We consider the interaction of these blobs with the magnetic field 
of a compact, accreting star, where diamagnetic blobs
suffer a drag. We show that when the magnetic moment is not
aligned with the spin axis, the resulting force is pulsed, and this
can lead to resonance with the oscillation of the blobs
around the equatorial plane; a resonance condition where energy is 
effectively pumped into non--equatorial motions is then derived. 
We show that the same resonance condition applies for the quadrupolar
component of the magnetic field. We discuss the conditions of applicability 
of this result, showing that they are quite wide. We also show that
realistic complications, such as chaotic magnetic fields, buoyancy, 
radiation pressure, evaporation, Kelvin--Helmholtz instability, and shear 
stresses due to differential rotation do not affect our results. In accreting 
neutron stars with millisecond periods, we show that this instability leads to 
Lense--Thirring precession of the blobs, and that damping by viscosity can
be neglected. 
\end{abstract}


\section{Introduction}

The interaction between the magnetic field anchored to rotating 
bodies and matter orbiting around them plays an important role in a variety of
astrophysical situations, ranging from the Jupiter-Io system 
(Stern and Ness 1982) to the Jovian ring (Burns \etal\/ 1985),
to protoplanetary disks around young magnetic stars (Bodenheimer 1995)
and accreting degenerate stellar remnants in binary systems, such as white 
dwarfs in intermediate polars or neutron stars in X-ray binaries (Frank,
King and Raine 1995). Straightforward applications of the laws of 
electrodynamics are possible in cases, such as the Jovian ring,
in which the orbiting matter is made of solid, dust particles. Due to a 
variety of poorly known MHD and plasma effects, modelling this interaction in 
the case of gaseous disks that orbit different classes of magnetic stars is far 
more  uncertain and difficult. In particular, theoretical descriptions of 
viscous accretion disks around magnetic rotating stars have been largely based 
on a number of simplifying assumptions. Two of these are especially relevant to 
the present work: (a) that the magnetic field (assumed dipolar) and rotation
axes of the star are coaligned;  (b) that the disk is continuous, smooth,  
azimuthally symmetric and lies in the equatorial plane of the rotating star.
For example Ghosh and Lamb (1979) adopted these assumptions and introduced 
an {\it ad hoc} effective diffusivity of the disk plasma to derive a 
stationary disk solution, in which the star magnetic field lines thread
the disk and slip across its material. Models of this kind proved very 
useful for evaluating the basic properties of the disk-magnetic field 
interaction, such as the size of the magnetosphere and the balance of the 
material and non-material torques. 

It has always been recognised that the coalignement of the magnetic field and
rotation axes of the star is an unrealistic hypothesis as a 
finite magnetic colatitude is required for the generation of the periodic
signals at the star spin frequency  that are often observed in these
systems. On the other hand, the pronounced aperiodic (or, in some cases, 
quasi-periodic) variability that is frequently detected on a variety of
timescales ranging from days to milliseconds in disk accreting compact stars
of all classes (intermediate polars: King and Lasota 1990; accreting neutron 
stars and black holes: van der Klis 1995, van der Klis 1997) testify that the 
disk cannot be regarded as smooth, continuous and azimuthally symmetric.
In a number of cases the quasi-periodic timing signature of blobs orbiting 
over a limited range of radii close to the compact star is
clearly seen. A highly inhomogeneous and clumpy disk, possibly comprising two
distinct and coexisting phases (hot and cold), is also envisaged as the end
product of the instabilities predicted by current models of viscous disks
(e.g the so-called secular and viscous instabilities of radiation-pressure
dominated $\alpha$-disks, Lightman and Eardley 1974, Shakura and Sunyaev 1976, 
or the magnetic amplification and buoyancy of flux tubes in dynamo-driven disks,
Vishniac and Diamond 1992), as discussed by Krolik (1998).

In all accreting objects endowed with a strong magnetic field, the
presence of inhomogeneities and/or blobs that can be regarded as 
discrete entities suggests that a novel mode of interaction
between the compact star and the disk is possible. This occurs because
individual blobs are most likely strongly diamagnetic (see for instance 
King 1993, Wynn and King 1995 and references therein); when moving through 
a magnetic field, strong surface currents develop on the blob the 
main effect of which is the generation of a drag opposite to the  component
of the blob velocity perpendicular to the field (Drell, Foley and 
Ruderman 1965). The acceleration acting on each blob is thus
\begin{equation}
\vec{a} = -\frac{\vec{v}_\perp^{(rel)}}{t_d}\;;
\end{equation}
the drag time--scale $t_d$ is given by
\begin{equation}
t_d = \frac{c_A m}{B^2 l^2}\;,
\end{equation}
where $c_A$ is the Alfv\`en speed in the magnetic field $B$, and $m$
and $l$ are the mass and characteristic radius of the blob. Here 
$\vec{v}^{(rel)}$ is the relative velocity between the blob and a magnetic
field line. It is
easy to see (Fig.1) that this acceleration has a component along
the disk axis which tends to lift the blob off the equatorial plane
where it is, at least initially, lying. It is the purpose of this paper 
to study this dynamical interaction, and to show how this alters
the conventional view of accretion disks onto magnetized compact stars.
In order to bring out the physical meaning of the instability most clearly,
we shall at first idealize these plasma blobs as point masses. 
In the next section, it will be shown that this interaction leads to
the lifting of blobs off the equatorial plane, at a resonance radius;
the conditions under which this result applies are discussed in 
Section 3. In Section 4, we relax the hypothesis that blobs are point masses,
and establish that a variety of effects, all related to the blobs having a 
finite size, do not modify our results. As the only concrete application
of this instability, we discuss in Section 5 the generation of modulation of 
the X--ray flux in Low Mass X--ray Binaries (LMXBs) exhibiting
millisecond QPOs at frequencies comparable to the single--particle
Lense--Thirring (1918) precession frequency. The last section summarizes 
the results.

\section{Dynamical analysis of the blobs' motion}

We begin by considering the motion of individual blobs initially lying
on an accretion disk around a magnetized compact object. The disk plane
is $z=0$, and the magnetic field is a pure dipole with magnetic moment
$\vec{\mu}$; 
at a point located at distance $R$ in a direction $\vec{n}$ from the star 
center, this is $\vec{B} = (3\vec{n}(\vec{n}\cdot\vec{\mu}) - \vec{\mu})/R^3$.
It is necessary that the dipolar moment not be aligned with
the star's rotation axis $z$; we shall take it to be inclined by an angle
$\beta$, so that its Cartesian components, in a non--rotating frame, are
\begin{equation}
\vec{\mu} = \mu_0 \left(
\begin{array}{c}
\sin\beta \sin \omega_s t \\
\sin\beta \cos \omega_s t \\
\cos\beta
\end{array} \right) \;.
\end{equation}
We have assumed here that the star rotates around the $z$~axis with angular
frequency $\omega_s$. The drag force of Eq. 1 can then be seen 
from Fig.1 to be oscillating, {\it i.e.}, the acceleration sometimes 
speeds up, sometimes decelerates the blob on its motion of revolution.
Thus the net effect of the drag force, except at resonances to be identified
in the following, is null: the overall structure of the accretion disk is
not changed by this effect. 

We assume here that the magnetic field is exclusively due to the accreting
object, neglecting the field distortion due to the disk, discussed by
Ghosh and Lamb (1979). We can easily see fom Fig. 1 that inclusion of
a $B_\phi$ component, which is parallel to the disk velocity, 
would leave the component of the  disk velocity perpendicular to the
magnetic field unaffected; thus the acceleration described in Eq. 1
would still have a component in the direction perpendicular to the disk
plane, leaving the analysis in the following, which will neglect
$B_\phi$, unaltered. 

We now idealize blobs as test particles, subject, beyond the usual
gravitational force, to the drag of Eq. 1 and to the usual viscous 
force $\vec{f}_V$. The equations of motion for each blob are then
\begin{equation}
\vec{\ddot{R}} = - \frac{GM}{R^3} \vec{R} - \frac{\vec{v}^{(rel)}_\perp}{t_d} 
+\vec{f}_V \;,
\end{equation}
where of course the blob speed perpendicular to the magnetic field is
\begin{equation}
\label{projection}
\vec{v}_\perp^{(rel)} = 
\vec{v}^{(rel)} - \frac{\vec{B} (\vec{B}\cdot\vec{v}^{(rel)})}{B^2}\;,
\end{equation}
and the relative velocity for circular orbits (the only ones we shall
be interested in, in the following) is given by
\begin{equation}
\vec{v}^{(rel)} = (v_K - \omega_s R) \hat{e}_\phi\;.
\end{equation}
Here $v_K = (GM/R)^{1/2}$ is the Keplerian velocity, and $\hat{e}_\phi$
a unit vector in the azimuthal direction.

We now assume that the usual viscous timescale $t_v$ is long, and the 
Keplerian period of rotation $t_K=2\pi/\omega_K$ around the compact object
short, compared to the drag time--scale, Eq. 2. In
other words, we shall assume the ordering 
\begin{equation}
\label{ordering}
t_K \ll t_d \ll t_v\;.
\end{equation}
The validity of this assumption will be discussed later, because it does not 
hold everywhere, but it will be shown to hold around the resonance radius 
which will be identified shortly. It should be noticed that the above ordering 
is that adopted by King (1993).

Because of this ordering, we shall neglect normal viscosity henceforth,
and use the small parameter $t_K/t_d$ to perform a first--order expansion.
In other words, we shall treat the motion as pure 
rotation in the equatorial plane ($z=0$) to order zero, and treat the 
off--plane motions as first--order perturbations. Then the drag force in
Eq. 1, which is much smaller than the gravitational force because we 
assumed $t_K \ll t_d$, will force motions in the $z$~direction.
The linearized equation of motion in the $z$~direction is then (Binney and
Tremaine 1987)
\begin{equation}
\label{eqnmotion}
\ddot{z} + \omega_K^2 z = -\frac{v_{\perp,z}^{(rel)}}{t_d}\;,
\end{equation}
and in computing the term $v_{\perp,z}^{(rel)}/t_d$ we use 
\begin{equation}
\label{unpert}
\vec{v}^{(rel)} = (v_K-\omega_s R) \left(
\begin{array}{c}
\sin(\omega_K t+\phi) \\
\cos(\omega_K t+\phi) \\
0 \end{array}\right)\;,\;
\vec{n} = \left(
\begin{array}{c}
-\cos(\omega_K t+\phi) \\
\sin(\omega_K t+\phi) \\
0 \end{array}\right)\;, 
\end{equation}
and $R$ constant. Here $\phi$ is the (soon to be discarded) phase of the 
orbital motion of the given blob.

With straightforward computations we find 
\begin{equation}
\vec{B}\cdot\vec{v}^{(rel)} = 
-\mu_0 (v_K-\omega_s R) \sin\beta \cos((\omega_K-\omega_s)t+\phi)/R^3\;,
\end{equation}
\begin{equation}
\vec{n}\cdot \vec{\mu} = \mu_0 \sin\beta \sin((\omega_K-\omega_s)t+\phi)\;,
\end{equation}
and then
\begin{equation}
\label{bsquared}
B^2 = \frac{3\sin^2\beta \sin^2[(\omega_K-\omega_s)t+\phi]+1}{R^6}
\mu_0^2
\end{equation}
\begin{equation}
v_{\perp,z}^{(rel)} = - (v_K-\omega_s R) 
\frac{\cos\beta \sin\beta \cos[(\omega_K-\omega_s)t + \phi]}
{1+3\sin^2\beta \sin^2[(\omega_K-\omega_s)t+\phi]}\;.
\end{equation}
The equation of motion, Eq. \ref{eqnmotion},  becomes 
\begin{equation}
\label{eqmotion2}
\ddot{z} + \omega_K^2 z = -\frac{v_K-\omega_s R}{t_d}
\frac{\cos\beta \sin\beta \cos[(\omega_K-\omega_s)t + \phi]}
{1+3\sin^2\beta \sin^2[(\omega_K-\omega_s)t+\phi]}\;.
\end{equation}

There is no need to make the above equation more explicit. In fact, it
can be seen through Eq. 2 and Eq. \ref{bsquared} that the term $t_d$ is
modulated at the frequency $\sin^2[(\omega_K-\omega_s)t + \phi]$, exactly 
like the denominator of the right--hand side of Eq. \ref{eqmotion2}. 
By suitably redefining the initial time we can drop the phase $\phi$.
Thus we see that the linearized equation of motion along the star spin
axis $z$ is that of a free oscillator plus a forcing modulated at the angular 
frequency $\omega_K-\omega_s$, which makes perfect sense once we move to
a non--inertial reference frame rotating with the blob. Also, it is 
necessary that the rotator be not--aligned, otherwise ($\beta=0$)
the right--hand side of Eq. \ref{eqmotion2} will vanish. It is interesting
to notice that this effect vanishes also for perpendicular 
rotators ($\beta=\pi/2$) because in that case the blob acceleration is
always in the equatorial plane.

The right--hand--side $h$ of Eq. \ref{eqmotion2} can then be expanded as
\begin{equation}
\label{h}
h = \cos[(\omega_K-\omega_s)t] f_0 (\sin^2[\omega_K-\omega_s)t]) = 
\Sigma_n a_n \cos[(2n+1)(\omega_K-\omega_s)t] \;.
\end{equation}
Clearly, there is a resonance whenever 
\begin{equation}
\label{minor}
\omega_K = \omega_s 
\end{equation}
\ie, at corotation. However, the corotation resonance is irrelevant: in fact, 
the term $v_K-\omega_s R$ in Eq. \ref{eqmotion2} vanishes at this resonance, 
showing that there is no forcing. This is reasonable: if there is no relative
motion between the blob and the magnetic field lines, the motional 
electric field $\vec{E} = \vec{v}^{(rel)}\wedge\vec{B}/c$ vanishes, and there 
will be no induced currents on the surface of the blob. Then the blob-field
interaction is reduced to the usual, much weaker coupling of diamagnetic
substances with the gradients of the field. We shall thus neglect the
corotation resonance $\omega_K = \omega_s$ henceforth.

The other resonances occur for $\omega_K = \pm (2n+1)(\omega_K-\omega_s)$, \ie\/
\begin{equation}
\label{resonanceall}
\omega_K = \left\{ \begin{array}{cc}
\frac{2n+1}{2n}\omega_s & \omega_K > \omega_s \\
\frac{2n+1}{2n+2}\omega_s & \omega_K < \omega_s 
\end{array}
\right.
\end{equation}
the upper branch corresponding to inner (with respect to corotation) resonances,
the lower one to outer resonances. All of these resonances are included
inside a thin radial annulus defined by
\begin{equation}
\label{resonance}
\frac{1}{2} \omega_s \leq \omega_K \leq \frac{3}{2} \omega_s\;.
\end{equation}
These resonances are often called secular: a free oscillator with natural 
angular frequency $\omega_K$ is forced with the same period. 
This is the sought--after resonance condition. When it is satisfied, the
forcing is in step with the free oscillation: it is upward when the 
restoring force is upward, and downward when the restoring force is downward, 
so that the total effect adds up, and is not averaged out. It should
be noticed that all resonance conditions do not depend on the exact value of
$t_d$, provided it is modulated through $B^2$ (Eq. \ref{bsquared}) only.

Analysis of the motion in the equatorial plane adds no new resonance: the
forcing $g$ is again provided by a term like $h$ in Eq. \ref{h}, and
the equations of motion for the small displacements are of the form
$\ddot{x} +\kappa^2 x = g$, where however the epicyclic frequency $\kappa$, for 
a $1/r$ potential, equals the Keplerian angular frequency $\kappa = \omega_K$
(Binney and Tremaine 1987). So, motion in the plane shows exactly the
same resonances as motion off the plane. The significance
of these resonances is that they will induce strong epicyclic motions 
around their unperturbed, circular orbits. Since however in a $1/r$ 
potential all orbits must close, the circular orbits become, after the
perturbation, ellipses. If the total number of blobs is small, they will
be unlikely to collide; if on the other hand, and contrary to evidence
from QPO observations, the total number is large, shocks will
develop and the problem will not be treatable in the point mass approximation
used here. 

\subsection{The quadrupolar field}

The relative strengths of higher--order multipoles in the expansion of compact 
stars' magnetic field are virtually unknown. For application to objects 
where the accretion disk is thought to extend to within a few stellar radii, 
(\eg, kilo--Hertz QPO sources), the quadrupolar component 
may be a substantial fraction of the overall field. It seems thus interesting 
to consider also the case where the field is entirely due to a quadrupole.
We show here that it produces no new resonances.

We consider a quadrupolar magnetic field, with a symmetry axis which
we take as the $z'$~axis, and again we take this axis to be inclined 
with respect to the star spin axis $z$. The more general case of a 
quadrupole magnetic field with no symmetry axis can be obtained by
a suitable superposition of two quadrupolar fields with distinct
symmetry axes. Since in the following analysis the $z'$~axis is definite 
but otherwise arbitrary, the general case introduces no new element, and the
conclusion that no new resonance is present, holds in this case as well.
The magnetic field is then
\begin{equation}
\label{quadrupole}
\vec{B}' = \frac{3 D}{R^5} \left(\begin{array}{c}
x' (1-5\frac{z'^2}{R^2}) \\
y' (1-5\frac{z'^2}{R^2}) \\
z' (3-5\frac{z'^2}{R^2})\end{array}
\right)\;.
\end{equation}
Here $D$ is eigenvalue of the tensor quadrupole moment along the symmetry
axis. 

The reference system $K'$ in which the quadrupole magnetic field has
the form shown above is carried around the star axis by its
rotation (Fig. 2). We call $K$ the inertial system of reference 
in which the stellar axis of rotation is called $z$, and the $x$ and
$y$ axes lie in the equatorial plane. In order to determine the rotation 
matrix ${\cal R}$ which transforms vector components from the $K'$
to the $K$ system of reference, we proceed as follows. Calling $\beta$
the angle between the $z'$ and $z$ axes, the components of the 
$z'$ axis in the $K$ frame are 
\begin{equation}
\label{zprime}
\hat{z}' = \left(
\begin{array}{c}
\sin\beta \sin\omega_s t \\
\sin\beta \cos\omega_s t  \\
\cos\beta
\end{array}\right)\;.
\end{equation}
Since the quadrupole field is symmetric about the $z'$~axis, we 
can choose the $x'$~axis to lie in the $xy$~plane; then, because
it must satisfy the conditions $\hat{x}'\cdot\hat{x}'=1$, and
$\hat{x}'\cdot\hat{z}'=0$, we find
\begin{equation}
\label{xprime}
\hat{x}' = \left(
\begin{array}{c}
\cos\omega_s t \\
-\sin\omega_s t \\
0
\end{array}\right)\;.
\end{equation}
Then the $y'$~axis is given by the cross--product
$\hat{z}'\wedge\hat{x}' = \hat{y}'$:
\begin{equation}
\label{yprime}
\hat{y}' = \left(
\begin{array}{c}
\cos\beta \sin\omega_s t \\
\cos\beta \cos\omega_s t \\
-\sin\beta
\end{array}\right)\;.
\end{equation}
Putting together Eqs. \ref{zprime}, \ref{xprime}, \ref{yprime}, we find for
the rotation matrix ${\cal R}$
\begin{equation}
\label{R}
{\cal R} = \left(
\begin{array}{ccc}
\cos\omega_s t & \cos\beta \sin\omega_s t & \sin\beta \sin\omega_s t \\
-\sin\omega_s t & \cos\beta \cos\omega_s t & \sin\beta \cos\omega_s t \\
0 & -\sin\beta & \cos\beta
\end{array}\right)
\end{equation}
and, as usual the inverse matrix which transforms vector components from
the $K$ to the $K'$ systems of reference is given by ${\cal R}^{-1} =
{\cal R}^T$. 

The magnetic field (Eq. \ref{quadrupole}) can be rewritten as
\begin{equation}
\vec{B}' = B_\circ \vec{r}' + \frac{6 D z'}{R^5} \hat{z}'
\end{equation}
and, by acting upon the first term, the rotation matrix still
returns a vector directed along $\vec{r}$. Since we need to evaluate 
the forcing term (Eq. \ref{projection}) for the unperturbed orbit
for which $\vec{v}\cdot\vec{r} = 0$, we see that the first term 
on the right--hand--side of the above equation disappears from both 
the term $\vec{B}\cdot\vec{v}$ and from the term $B_z$, so that,
except for the term $B^2$ in Eq. \ref{projection}, we need consider only
the term $\frac{6 D z'}{R^5} \hat{z}'$.

We now take for our unperturbed orbit Eq. \ref{unpert}.
Acting with ${\cal R}^T$ upon $\vec{r}$ we find
\begin{equation}
\label{zprime2}
z' =  R \sin\beta \sin[(\omega_K-\omega_s)t+\phi]
\end{equation}
and from this, transforming $\vec{B}'$ with ${\cal R}$, we find
\begin{equation}
\label{bnotprime}
\vec{B} = \frac{6 D z'}{R^5} \left(
\begin{array}{c}
\sin\beta \sin\omega_s t \\
\sin\beta \cos\omega_s t \\
\cos\beta
\end{array}\right)\;.
\end{equation}
From the above we now find
\begin{equation}
\vec{B}\cdot\vec{v} = \frac{3 D (v_K-\omega_s R)}{R^4} \sin^2\beta
 \sin[2(\omega_K -\omega_s)t + \phi]
\end{equation}
and
\begin{equation}
B_z = \frac{6 D}{R^4} \sin\beta \cos\beta \sin[(\omega_K-\omega_s)t + \phi]\;.
\end{equation}
Of course $B^2$ is invariant by rotation, so it is most easily computed
in the $K'$ frame; from Eq. \ref{quadrupole} we have
\begin{equation}
B^2 = \frac{9D^2}{R^8} (1+5\frac{z'^4}{R^4}-2\frac{z'^2}{R^2})\;,
\end{equation}
with $z'$ now given by Eq. \ref{zprime2}.
Inserting this into Eq. \ref{projection} we have
\begin{equation}
v_{\perp,z}^{(rel)} = (v_K-\omega_s R) \frac{\sin^3\beta \cos\beta 
\sin[(\omega_K-\omega_s)t+\phi]
\sin[2(\omega_K-\omega_s)t+\phi]}{1+5\frac{z'^4}{R^4}-2\frac{z'^2}{R^2}}\;.
\end{equation}

Proceeding as after Eq. \ref{eqmotion2}, we see that the forcing term
can be developed in terms of the form $\cos[(2n+1)(\omega_K-\omega_s)t]$;
we thus obtain the same corotation resonance as in 
Eq. \ref{minor}, which again can be neglected for the same reasons 
discussed in the previous section, 
and the other resonances given by $\omega_K = \pm (2n+1)(\omega_K-\omega_s)$, 
again identical to the previous section. Since we treated the problem in the 
linear regime, the sum of the dipole and quadrupole fields also shows
the same resonances. Once again, motion in the equatorial
plane adds nothing new.

\section{Conditions of applicability}

We cannot compute an accurate value for $t_d$, for several reasons.
First, we are not even close to knowing the blobs' properties (but 
see the next section) so that
the quantity $m/l^2$ which appears in Eq. 2 (\ie, the blobs' typical
surface density) is basically unknown. We shall take $m/l^2 = q
\Sigma$, with $\Sigma$ the unperturbed disk's surface density at the same 
radius, and expect $q >1$, because blobs must form from the contraction of disk 
material. Second, when blobs form, the interblob space is filled with tenuous 
gas of unknown density, which clearly makes $c_A$ larger than it would be should
there be no blobs. Thus, in order to establish the second half $t_d \ll t_v$ of
the ordering in Eq. \ref{ordering}, we take for $c_A$ the speed of light
$c$. We now have
\begin{equation}
t_d \la \frac{q c \Sigma}{B^2}\;.
\end{equation}

The unperturbed disk's viscous inflow time--scale $t_v = R/v_r = 
2\pi R^2\Sigma/\dot{M}$. We shall use this unperturbed value, despite
the fact that this is most likely a lower limit to the time--scale,
because, if many large blobs form, the interblob disk will be filled
with lower density gas, which obviously exerts a weaker viscous drag
on the blobs. Using the definition of Alfv\`en radius (Shapiro and
Teukolsky 1983), and the fact that 
the magnetic field is dipolar $B= B_\circ (R_{S}/R)^3$, we obtain
\begin{equation}
\frac{t_v}{t_d} \ga \frac{2^{3/2}\pi }{q}
\left(\frac{1}{\alpha}\right)
\left(\frac{R}{R_A}\right)^{1/2}\;.
\end{equation}
We find that $t_v/t_d \ga 1$ provided $q \alpha\la 10$. 
Here $\alpha< 1$ is the Shakura--Sunyaev adimensional viscosity parameter.
Despite several serious uncertainties, it appears the key assumption
$t_v\gg t_d$ must hold at least in the case of relatively small values of
$\alpha$. We feel we cannot usefully speculate more about the value of $q$.

In order to establish the second half of the ordering, $t_K \ll t_d$, 
we take a lower limit to $t_d$, by using all quantities as if the disk
were totally unperturbed. In the innermost, radiation--pressure 
dominated region of accretion disks (region A of Shakura and Sunyaev 1973),
we obtain
\begin{equation}
\frac{t_d}{t_K} = 12 \alpha^{-3/2}
m^{-3/2} \dot{m}_{16}^{-2}
\left(\frac{R}{R_{S}}\right)^9\;,
\end{equation}
where $m$ is the mass of the accreting object in solar units and $\dot{m}_{16}$
the accretion rate in units of $10^{16} \;g\;s^{-1}$. 
In region B the inequality $t_d/t_K \gg 1$ is always satisfied.
Here $R_{S}$ is the star radius; since typcally $R \ga 2 R_{S}$,
the lower limit being dictated by the millisecond QPOs of some accreting
neutron stars, the ordering $t_d \gg t_K$ is satisfied.
The further condition, that the blob stays on the resonance long enough 
for significant energy transfer to occur, before the viscosity pushes it
inward and away from the resonance, is satisfied provided $t_d \ll t_v$,
and thus no distinct requirement is added. 
The requirement that the outermost resonance exists is of course that the 
Alfv\`en radius occurs inside the resonance radius, or, equivalently,
that the magnetic star is a sufficiently slow rotator (Ghosh and Lamb 1979).
Calling $\omega_A$ the Keplerian angular velocity of matter at the Alfv\`en 
radius, this condition
is $\omega_A \geq \omega_s/2$. This is less stringent than the
usual condition $\omega_A \geq \omega_s$, which implies that matter 
that becomes attached to the stellar magnetic field lines does not get
flung out by the propeller effect (Illarionov and Sunyaev 1975). 

\section{On the finite size of the blobs}

In the previous sections we gave a highly idealized account of the resonance
mechanism. This was done partly to obtain a tractable problem, and partly
to bring out clearly the meaning of the instability. In this section we
reintroduce some of the complications that were left out, in particular
random magnetic fields, buoyancy, radiation pressure and blob heating. All of 
these effects will be negligible in the limit of very small blob size; while
this is quite clear for buoyancy, radiation pressure and heating, it is less so
in the case of random magnetic field. So we first discuss this last
point, then turn to the others to deduce upper limits to the blob size.
It is convenient to introduce for blobs an `equivalent length' $l_e$, 
distinct from the physical length $l$, and defined as
\begin{equation}
\label{equivalent}
l_e^2 \equiv \frac{m}{\Sigma}\;.
\end{equation}
$l_e/l$ is essentially the factor of contraction of the blob material,
and $(l_e/l)^2$ the density contrast with respect to the surroundings. 

\subsection{Chaotic magnetic fields in generic accretors\label{zumpa}}

From the discussion in the Section 2, it seems likely that,
as long as the magnetic field is an ordered one, the structure of
the resonances is the one described in Eq. \ref{resonance}. However,
outside the magnetosphere, the magnetic field is likely to have also a chaotic 
component, the main sources of which are all the irregularities of the disk,
foremost among which are the very blobs we discussed. In
fact, the diamagnetic blobs have surface currents which themselves 
generate a magnetic field. As seen from one of the blobs, the total
magnetic field will thus be chaotic, with a Fourier spectrum (with respect to
time) strongly peaked around the frequency $\nu = 0$. The reason is that
the magnetic field will appear as the incoherent superposition of
randomly oriented magnetic moments (corresponding to the independent
blobs), and since dipole magnetic fields scale with distance from the
magnetic moment as $R^{-3}$, the local magnetic field will be strongly
dominated by the nearby blobs, which nearly corotate with the
blob we are considering, leading to a spectrum strongly peaked around
$\nu = 0$. 

The cut--off frequency $\nu_c$ at which the chaotic magnetic field power 
spectrum starts to decrease is difficult to estimate. It is clearly of
order $\nu_c \approx \nu_1 - \nu_b$, where $\nu_b$ is the Keplerian 
frequency at the radius $R$ under consideration, and $\nu_1$ is the 
Keplerian frequency at the radius $R_1$ such that the total number of blobs 
in a circle centered on $R$ and of radius $R_1 - R$ is a few. Beyond this 
radius, and thus this frequency difference, the magnetic moments begin to 
add incoherently and thus to cancel each other. However, due to our ignorance
of blobs' properties, including number and distribution, we are unable
to estimate $\nu_c$. 

Despite the near--cancellation of the many, distant and incoherent moments, a 
small residual field (the `square root effect') will survive at all frequencies,
but its dynamical effect at 
frequencies different from the Kepler frequency is negligible: we showed
in fact that these components of the magnetic field cause oscillating
forces out of step with the natural oscillation frequency, and thus a null 
average effect. Only the component of the chaotic field at the oscillation
frequency could be important. Consider then a chaotic magnetic field 
$B_c$ with (time) Fourier spectrum ${\cal B}(\nu)$, as seen by the orbiting 
blob; if the resonance width is $\delta\!\nu$, then the relevant component
of the chaotic magnetic field is given by ${\cal B}(\nu_K) \delta\!
\nu$. The condition under which the analysis of the previous section
applies is then 
\begin{equation}
\label{chaotic}
{\cal B}(\nu_K) \nu_K \frac{\delta\!\nu}{\nu_K} \ll B_o
\end{equation}
where $B_o$ represents the ordered magnetic field. This is much less 
restrictive than the naive condition $B_c \ll B_o$, for two reasons. 
First, as discussed above, most of the power of the chaotic magnetic 
field will be concentrated around zero frequency, so that 
${\cal B}(\nu_K) \nu_K \ll B_c$, and second, for 
blob of angular size $l \ll l_e$, $\frac{\delta\!\nu}{\nu_K}
\approx l/l_e \ll 1$. In the next subsection we shall establish a 
quantitative upper limit to the small--scale chaotic magnetic field in
a restricted but important subclass of accretors. 

\subsection{Chaotic magnetic fields in neutron stars with millisecond QPOs}

While the analysis of the previous paragraph holds for all accretors, 
in this specific class of accretors (van der Klis 1997) a stronger argument
can be made which rules out the importance of chaotic magnetic fields. The
energy density of the neutron star magnetic field at a given radius can 
be compared to the local, internal energy density $\epsilon_{th} = 
3 \rho_d k T/2 m_p$, where
$\rho_d$ is the disk average density, $k$ Boltzmann's constant, and $m_p$
the proton mass. In fact, since the local magnetic field must be generated
at the expense of the disk's internal energy, this provides an absolute
upper limit to the chaotic field energy density. 

The dipole ($B_d$) energy density at radius $R$ can be rewritten as 
\begin{equation}
\label{dipoledensity}
\frac{B_d^2}{8\pi} \approx \rho_A v^2_A \left(\frac{R_A}{R}\right)^6
\end{equation}
where $\rho_A$, $v_A$ are the disk density and rotational velocity
at the Alfv\'en radius, $R_A$. This is because, by definition, the magnetic
and rotational energy densities are roughly matched there. On the other hand,
the chaotic field $B_c$ energy density obeys
\begin{equation}
\label{chaoticdensity}
\frac{B_c^2}{8\pi} \la \epsilon_{th} = \frac{3 \rho_d kT}{2 m_p} 
\end{equation}
and this can be rewritten, taking into account the fact that, in the case 
of such fast and low--magnetic field accretors the Alfv\'en radius surely
falls into region A of the accretion disk (Shakura and Sunyaev 1973), as
\begin{equation}
\label{better}
\epsilon_{th} = \frac{3 \rho_A k T_A}{2 m_p} \left(\frac{R}{R_A}\right)^{9/8}\;.
\end{equation}
Taking the ratio of Eq. \ref{dipoledensity} and \ref{better} we find
\begin{equation}
\label{ratio1}
\frac{B_d^2}{B_c^2} = \frac{2 m_p v_A^2}{3 k T_A} 
\left(\frac{R}{R_A}\right)^{-57/8}\;;
\end{equation}
and again using the expressions for region A we find
\begin{equation}
\label{aux}
\frac{2 m_p v_A^2}{3 k T_A}  = 9.4\times 10^5 m^{25/24} \alpha^{1/4} 
\left(\frac{\nu_A}{10^3\; Hz}\right)^{5/12}
\end{equation}
where we used the frequency at the Afv\'en radius $\nu_A$ to eliminate the 
Alfv\'en radius from the above expression. 
It is convenient to rewrite Equation \ref{ratio1} with the aid of Eq. \ref{aux}
and in terms of Keplerian frequencies; defining $\omega_{e}$ as the frequency 
where the magnetic energy densitites are equal, we find
\begin{equation}
\label{ratio2}
\nu_e = 18\; Hz \; m^{-25/82} \alpha^{-3/41} \left(\frac{\nu_A}{10^3\; Hz}
\right)^{36/41}\;.
\end{equation}
In neutron stars exhibiting millisecond QPOs, we have $\nu_A \approx 1000\; Hz$ 
and the pulsar spin frequency $\nu_s \approx 300 \; Hz$ (van der Klis 1997). 
The outermost resonance ($\omega_K = \omega_s/2$, Eq. \ref{resonance})
occurs for $\nu_K/\nu_A \approx 0.15$ for these values.
Thus the dipole magnetic field exceeds the internal energy density of the
disk for all radii for which the Keplerian frequency exceeds $\nu_e$,
and this includes in particular the whole resonance strip, Eq. \ref{resonance},
especially when one remembers that this is an upper limit to the chaotic 
magnetic field, and the arguments of the previous paragraph. 

The above analysis would be flawed only in the case in which the energy density 
of the chaotic magnetic field were in rough equipartition with the energy 
density of the radiation, which dominates the energy balance in region A. 
However, this seems somewhat less likely, given that proccesses leading
to field amplification (shearing, compression, turbulent dynamo) couple 
the magnetic field preferentially to the matter. 

\subsection{Buoyancy}

Buoyancy tends to reduce the restoring force, in our case gravity.
Calling $\rho_b$ and the $\rho_d$ the
densities of the blob and of the disk (this is a local quantity), 
buoyancy reduces the local gravitational acceleration by the factor
$z = (\rho_b-\rho_d)/\rho_b$; it enters our Eq. \ref{eqnmotion}
by reducing the square of the Kepler frequency by the same factor. 
By neglecting buoyancy, we have essentially assumed $\rho_b \gg 
\rho_d$, which implies that the correction factor $z \approx 1$, so
we acted consistently with our approximation of point--mass blobs.
When the density contrast of the blobs becomes small, the restoring
force is greatly weakened, the restoring period (previously $t_K$, now
$\rho_d/(\rho_b-\rho_d) t_K \gg t_K$) will exceed $t_d$, and the 
approximations leading to Eq. \ref{eqnmotion} will fail: in this case
no resonance is possible, because the restoring period is pushed to
infinity. But then, the condition of applicability of our analysis is
that the density contrast $(\rho_b-\rho_d)/\rho_d \ga 1$, \ie\/
$l/l_e \la 1/2$, roughly as determined at the end of Section \ref{zumpa}.

\subsection{Shear stresses}

Realistic blobs will be subject to severe shear stresses due to differential 
rotation, which will elongate them in the azimuthal direction. However, the 
final consequences of this shear are not very important. In fact, the number of 
orbital revolutions that the blobs can execute before differential rotation 
smears them over, say, an azimuth of 1 radian (\ie, before axial symmetry is 
restored) is given by $N \approx \omega_K/\delta\!\omega_K$, where
$\delta\!\omega_K$ is the variation of the Kepler frequency across the blob;
using our hunch that typical blobs sizes are $\approx H$, the disk 
semi--thickness, we find $N \approx \frac{R}{H} \gg 1$. In particular, in
region A of viscous accretions disks (Shakura and Sunyaev 1973), where the
semi--thickness is given by (neglecting the disk tapering at its inner edge)
\begin{equation}
\label{ha}
H = 10^4\; cm \; \dot{m}_{16} 
\end{equation}
we obtain
\begin{equation}
\label{lifetime}
N \approx 510 m^{1/3} \dot{m}_{16}^{-1} 
\left(\frac{10^3\;Hz}{\omega_K}\right)^{1/3}\;.
\end{equation}
During the stretching, blobs evolve at roughly constant volume (and hence
constant density), otherwise the declining internal pressure would be overcome
by the outer, confining pressure. Thus stresses generate a filament which
will still be subject to the magnetic drag instability discussed here. 

In the following Section \ref{lt}, we shall apply this instability to the 
generation of Lense--Thirring precession around Atoll sources exhibiting 
millisecond QPOs. Here, the blob lifetime $t_b$ has a lower limit given by the
observed width of the line due to LT--precession ($\delta\!\nu 
\approx 10\; Hz$, Stella and Vietri 1998), which is to be compared with the 
Kepler frequency at the same radius, $\nu_K \approx 10^3\; Hz$. We have
$t_b \ga \nu_K /\delta\!\nu \approx 100$, consistent with the estimate 
of Eq. \ref{lifetime}. 

What this amounts to is that we should see our picture as applied to a bead 
of blobs rather than a single one, an inessential modification. Thus shear 
stresses are unimportant, provided of course fresh, tiny blobs ($l \approx H$)
are injected continuously at the resonances. 

\subsection{Radiation pressure}

Once the blobs are lifted by the instability off the disk, they will exit
the disk shadow and be exposed to radiation pressure from the accreting source.
Clearly, if the radiation from the compact object is beamed, this force will 
be pulsed at frequency $\omega_K-\omega_s$, but it will also have a non--zero 
average: radiation pressure is outward. If this force on the blobs 
exceeds the magnetic drag (either in or out of the plane), our previous analysis
will fail. The importance of radiation pressure is of course dependent
upon the nature of the sources, being larger in neutron stars than in
white dwarfs, and in particular in Z--type LMXBs more than in Atoll sources.

We consider an accretor with luminosity $L$ a fraction $f$ of Eddington's
($L = f L_E$, with $ f < 1$). The radiation force is $F_{rp} = 
L l^2/4 c r^2$, and the gravitational force $GM m_b/r^2$ with $M$ the mass of 
the accretor and $m_b$ the blob mass; the ratio ${\cal W}$ between the two is
\begin{equation}
{\cal W} = \frac{L l^2}{4 G c M m_b}\;,
\end{equation}
and, using the definition of $L_E$, and our expectation that the blobs
will have typical sizes of order $l \approx H$, the disk semi--thickness,
\begin{equation}
{\cal W} = f \frac{\pi m_p H^2}{m_b \sigma_T} = \frac{f}{\tau_{es}}
\end{equation}
where $\sigma_T$ is the Thomson cross--section, and $\tau_{es}$ is the optical 
depth to Thomson scattering of the blob. Since we took blobs to be at least
a few times denser than the disk at the same location, the optical depth of the 
blob is a few times that of the disk in the direction perpendicular to the 
plane. In conventional viscous accretion disks $\tau_{es}$ is a monotonically 
increasing function of radius; thus, by considering Z--type sources (which are,
together with Atoll sources, the fastest rotators, but have luminosities larger 
by about a factor of 100) we are putting ourselves into the worst case. Even 
then, $\tau_{es} \approx 200-700$ at the radii corresponding to Eq. 
\ref{resonance}, so that we find at those radii
\begin{equation}
{\cal W} \la 6\times 10^{-4} \frac{f}{0.3} \frac{500}{\tau_{es}}\;.
\end{equation}
Thus we have established that, at such large radii, radiation pressure is
negligible in Z--type sources and, {\it a fortiori}, in Atoll sources for which 
$f \approx 0.01$. This conclusion is further strengthened by the remark that in
the above we assumed the luminosity to be spherically symmetric, while, if
we assume some beaming as is most natural in spinning neutron stars, the
relevant luminosity to be used above is smaller than the value employed. 
Also, the conclusion holds also for $\beta$--disks (Stella and Rosner 1984). 

\subsection{Blob heating}

A more subtle effect, the removal of angular momentum from an orbiting object, 
is discussed by Miller and Lamb (1993, 1996), and more recently by Miller 
(1998). This is an interesting effect which, when coupled with the excitation
mechanism discussed here, might lead to blobs' precession around the neutron
star. We discuss here two objections to Miller's (1998) computations, one 
applying in the case in which blobs are evaporated quickly by the neutron star 
luminosity, the other one in the opposite limit of no evaporation. Taken 
together, these two arguments show that the importance of angular momentum
removal by radiation pressure forces is still too rough to be of great cogency.

The importance of evaporative effects depends in a sensitive way upon such 
unknown quantities as the blobs' cross--section to the neutron star, its 
chemical composition, conduction and its inhibition by small--scale tangled 
magnetic fields, and so on. So we shall simply consider the implications of
two limiting cases, quick or slow evaporation, without consideration of
which one actually holds. 

In the case of quick evaporation, a diffuse layer of matter will be produced 
by the evaporation of the first 
blobs to be lifted off the disk, in which following blobs will be protected 
from sunburns. The physical conditions of this layer remind us of the so--called
hot inner disk corona (Boyle, Fabian and Guilbert 1986, and especially
Kallman and White 1989). Thus, this may be thought of as 
one of the mechanisms providing the material of the corona, with the process 
continuing until saturation: when the hot corona becomes so optically thick 
in the radial direction to shield new blobs, it stops accreting new matter, 
in a self--regulating way. It should furthermore be remarked that the fact that
the hot disk corona is optically thick in the radial direction by no means
implies that it is optically thick also in the direction perpendicular to the 
disk. Thus the resulting optical depth along a random line of sight
may very well be smaller than $1$, thus leaving observational properties
unaffected. Also, the hot disk corona forms in rough pressure equilibrium with 
the disk, ensuring the pressure confinement of the blobs. 

In the opposite limit, in which we totally neglect evaporation, the following 
argument shows that the computations by Miller (1998) are not relevant to the 
problem at hand. The width of the kHz QPO peak ($\approx 10\; Hz$) indicates 
a blob lifetime of at least $0.1\;s$ and a radial extent of $\approx 10^4\; cm$ 
or less (cf. Miller, Lamb and Psaltis 1998), in agreement with our hunch
that a typical blob raidus equals the disk thickness, Eq. \ref{ha}. 
If the rms amplitude of the QPOs, often in the few percent range, is indicative 
of the gravitational energy released by the blobs, then at least some $10^{34}\;
erg\; s^{-1}$ are produced by the blobs in an Atoll LMXRB at any time.
For the kHz QPO signal not to be smeared by the superposition of
signals  from different blobs with different phases, the number 
of blobs that are active at any time must be limited to, say, 
$\sim 10$. For an energy conversion efficiency of 0.1, the 
numbers above convert to a blob mass of $\sim 10^{14}\; g$ and a (radial)
Thomson  depth, $\tau_b$, in the $10^5$ range. 
The radiation drag discussed by Miller (1998) operates only within a few
optical depths (say 3), such that only the blob outer layers facing the 
neutron star  are affected by radiation drag at any time. This results in a 
gradual ablation of the blob, on a timescale of $\tau_b \Delta \varphi / 
3\Delta \omega \geq 3\; s$, where $\Delta \varphi \geq 10^{-2}\; rad$ is the 
azimuthal extent of the blob and $\Delta \omega \sim 10^2\;rad\;s^{-1}$ 
the change of angular velocity caused by radiation drag (cf. Miller 1998). 
This radiation drag timescale is far longer than the blob lifetime
estimated from the QPO peak widths. We therefore conclude, at variance with 
Miller (1998), that the motion of the blobs (including their Lense-Thirring 
precession) is virtually unaffected by radiation drag, even in the case in 
which there is no accretion disk corona that shields the blobs from the 
neutron star radiation. 

\subsection{Finite size}

A finite size may have another interesting consequence: blobs may straddle 
several resonances simultaneously, especially close to corotation which, 
according to Eq. \ref{resonanceall}, is an accumulation point of resonances. It 
is not clear what will happen to the blobs in that case, because the 
forces which assure the blob cohesion, namely pressure confinement and
especially the large internal viscous stresses, may well overcome the forcing 
due to several, out of phase resonances. It seems indeed more likely that only 
the two extreme resonances, $\omega_K = \omega_s/2$ and $\omega_K = 3/2 
\omega_s$ may be capable of lifting blobs off the equatorial plane. The
reason for this is that, in a $1/r$ potential the radial separation
$\delta\! r$ of resonances obeys $\delta\! r/r \approx 1$ only for these 
two, outermost resonances. Then, since we showed that the condition for
the correctness of the analysis in the previous sections is $l/l_e \la 1/2$,
we deduce that only these two resonances are sure to be effective. More
resonances may be active in the more extreme case $l/l_e \ll 1$. It should
also be stated that we do not expect this instability to clear out the disk
at resonances, because the diffuse component of the disk will not be 
subject to the magnetic drag, and will continue to accrete onto the
central object as if unperturbed. 

\subsection{Kelvin--Helmholtz instability}

One of the side--effects of the motion of the blobs with respect to the
average, diffuse disk matter is the appearance of the ever--present, 
ever--destructive Kelvin--Helmholtz instability ($\equiv$ KHI). However, its 
development is surely slow compared to the development of the instability 
considered here. The reason is that the KHI growth rate $\Gamma$ is
\begin{equation}
\label{khi}
\Gamma = \frac{2\pi}{l} V_z \left(\frac{\rho_d}{\rho_b}\right)^{1/2}\;.
\end{equation}
Here $V_z$ is the off--plane velocity induced by the magnetic drag force,
and as usual we suppose $l \approx H$. However the growth rate for
the instability discussed in this paper is $V_z/H$, which exceeds 
$\Gamma$ by the factor $(\rho_d/\rho_b)^{1/2} \ll 1$ by assumption. When
the blob is lifted off the equatorial plane, it will presumably find
itself into an even more tenuous medium, and the KHI will slow down
further still. Furthermore, the KHI can be tamed in a suitably radiative
medium (Vietri, Ferrara and Miniati 1997), so that it seems at least
likely that it will not be a major source of destruction for the blobs.

This Section is well--summarized then by stating that all of the limiting 
factors discussed here are negligible, provided the density contrast of
blobs orbiting a generic accretor exceeds unity, $(\rho_b-\rho_d)/\rho_d
\ga 1$, or, equivalently, $l/l_e \la 1/2$. 

\subsection{On the origin and later evolution of the blobs}

Dishomogeneities in the accretion flow may arise as a consequence of either
disk--specific instabilities (Lightman and Eardley 1974, Shakura and Sunyaev
1976, Vishniac and Diamond 1992), or of the more conventional thermal 
instability (Field 1965). Until now, theoretical studies have not concentrated 
on the possible non--linear fate of local instabilities (while there is an 
extensive literature on the non--linear development of collective instabilities)
and, as a consequence, no credible estimates of either linear size or masses 
exist to date in the literature. The only remark we can make is that lack of 
theoretical
arguments is not lack of dishomogeneities, as testified by the panoply of QPOs
being discovered in a wide variety of accreting sources (intermediate polars: 
King and Lasota 1990; accreting neutron stars and black holes: van der Klis 
1995, van der Klis 1997). This wide variety points to the existence of a
robust mechanism leading to the formation of blobs, irrespective of the
nature of the accreting source. 

It is also clear that in the subsequent, non--linear development of this 
instability, hydrodynamical as well as dynamical effects will play 
significant roles. In particular, important effects which may be expected
to play a role are a drag by the disk material on the blobs as they 
plunge through the accretion disk, the possible formation of shocks if
the vertical speeds exceed the local sound speed as may well be expected to
be the case, or the excitation of spiral waves in the disk surface density
distribution. Hydrodynamic phenomena will limit the growth of the
instability discussed in this paper, but the important question of whether 
they manage to do so before individual blobs are lifted off the plane 
will need to be settled through full numerical simulations which lie outside the
scope of this paper. 

\section{Observational consequences\label{lt}}

We have until now, avoided the question of which, if any, obsevational
consequences may arise from the proposed instability. The most remarkable one 
is, most likely, the low--frequency modulation of the X--ray flux from the 
compact object around which the blobs are orbiting, with frequencies of order 
of the single--particle Lense--Thirring ($\equiv LT$) precession frequency. 

Lense--Thirring precession (Lense and Thirring 1918) by point masses
around compact, rotating objects in General Relativity is a slow precession
of the orbital plane induced by the compact object rotation, which can occur
provided the point mass does not lie in the rotating object's equatorial
plane. We showed (Stella and Vietri 1998) that the peak frequencies of some
spectral features in Low Mass X--ray Binaries exhibiting millisecond QPOs
are well--explained by the frequencies of Lense--Thirring 
precession, including their variation with source intensity. 

This effect may arise as follows. If the accretion disks around these
sources are indeed blobby, the blobs are lifted off the neutron star
equatorial plane by the instability discussed here, and then they 
LT--precess at the single--particle precession frequency. As discussed
above, we believe radiation pressure forces to be negligible. Then, 
if the blobs are self--luminous, or if they occult the
emitting areas on the surface of the neutron star, the resulting 
X--ray flux will be modulated also at the single--particle LT frequency.
Alternatively, the blobs may punch holes in the accretion disks at the two 
points where the inclined blob orbit intersects the disk. In this
case, the X--ray flux from the accreting disk itself will be modulated
at twice the single--particle precession frequency. 

Lense--Thirring precession of material from viscous accretion disks has been 
until now discounted because a large body of literature (Bardeen and 
Petterson 1975, Hatchett, Begelman and Sarazin 1981, Papaloizou and Pringle 
1983) has shown that viscosity manages to bring
the disk material to the neutron star equatorial plane, even if it lies
initially in a plane tilted with respect to it. We accept these conclusions, but
wish to show what follows: that the time--scale for viscous damping of 
off--plane motions in the disks' {\bf innermost} regions is long compared to the
time--scale on which LT--precession is forced. We would like to do this for a 
reasonable value of the viscosity, but since we are unable to specify which 
fraction of the disk material is in blobs and which is in the tenuous interblob 
medium, we do not know the viscosity. Thus we shall assume that in the 
z--direction viscosity is as large as in the tangential direction (an upper 
limit), and furthermore that the viscosity of the unperturbed Shakura and 
Sunyaev (1973) models holds (another upper limit). So the conclusion that 
viscosity is unable, at small radii, to damp LT--precession holds {\it a 
fortiori} for blobs in a realistic medium. 

The above argument can also be stated otherwise. We shall show that
the reason why viscosity is so effective at damping LT--precession is
that it does so at large radii, where viscous time--scales are much
shorter than LT--ones; where the opposite is true, \ie, at small radii,
the only reason why the viscous accretion disk lies in the equatorial plane
is that the material it is made of was obliged to lie down at large 
radii. Should any perturbing agent, such as the one discussed in this paper,
lift material off the equatorial plane, then at small radii we cannot count on 
viscosity to make the disk (and thus also the blobs) lie down once again. 

\subsection{On the dynamics of Lense--Thirring precession}

We consider a tilted, warped disk, split into azimuthally symmetric, coplanar
ringlets each identified by a local normal unit vector $\vec{l}$ which, for
small tilt angles, can be written as
\begin{equation}
\vec{l} \approx (\beta\cos\gamma, \beta\sin\gamma, 1)\;,
\end{equation}
with $\beta$ the tilt and $\gamma$ the twist. It is convenient, following
Hatchett, Begelman and Sarazin (1981), to introduce the complex quantity
\begin{equation}
W \equiv  \beta\cos\gamma +\imath\beta\sin\gamma
\end{equation}
which obeys the equation (Papaloizou and Pringle 1983)
\begin{equation}
\label{master}
\frac{\partial W}{\partial t} + (V_R + \frac{3\nu_1}{2 R})\frac{\partial W}
{\partial R} - \imath \omega_p W = 
\frac{1}{2\Sigma R^3 \omega_K}\frac{\partial}
{\partial R}(\nu_2\Sigma R^3\omega_K \frac{\partial W}{\partial R}) \;.
\end{equation}
The right--hand--side is a diffusive term which describes the damping 
by viscosity, while the rightmost term on the left--hand--side
describes the forcing due to the LT--precession. Here $V_R$, $\nu_1$ and 
$\nu_2$, are the radial velocity and $R-\phi$ and $R-z$ kinematic viscosities 
in the unperturbed disk. The Lense--Thirring precession, due to a compact 
object of mass $M$ and angular momentum $J = I \omega_s$,  acts on a timescale 
$t_p = \omega_p^{-1}$, where
\begin{equation}
\label{omegalt}
\omega_p = \frac{2 G J}{R^3 c^2}\;.
\end{equation}

We want to compare $\omega_p^{-1}$ with the timescale $t_v$ on which
viscosity leads to a damping of the tilt. We thus approximate the
viscosity--induced diffusive term on the right hand side of Eq. \ref{master}
as
\begin{equation}
\label{approx}
\frac{1}{2\Sigma R^3 \omega_K}\frac{\partial}
{\partial R}(\nu_2 \Sigma R^3\omega_K  \frac{\partial W}{\partial R}) \approx
\nu_2 \frac{W}{2 R^2} \equiv \frac{W}{t_v}
\end{equation}
from which we find 
\begin{equation}
\label{tv}
t_v = \frac{2 R^2}{\nu_2}
\end{equation}

Next we derive the adimensional ratio
\begin{equation}
\label{ratio}
\frac{t_v}{t_p} = t_v \omega_p = \frac{2 R^2}{\nu_2} \frac{2 G J}{R^3 c^2}\;.
\end{equation}

A useful modification to the theory given by Papaloizou and Pringle (1983)
has been introduced by Markovi\'c and Lamb (1998), who generalized the
above equation to the case of non--isothermal disks; we shall follow here their 
treatment. They write the $R-z$ viscosity as $\nu_2 \equiv \kappa \nu_1$,
where $\nu_1$ is the $R-\phi$ viscosity; they write, for a disk 
with a power--law temperature profile given by $T = T_i (R/R_i)^{-\mu}$, as
\begin{equation}
\nu_2 = \frac{4\alpha\kappa k T_i}{3m_p (GM)^{1/2}} R_i^\mu R^{3/2-\mu}\;.
\end{equation}
Introducing this into Eq. \ref{ratio}, we find
\begin{equation}
\label{ratio3}
\frac{t_v}{t_p} = \left( \frac{R_c}{R} \right)^{5/2-\mu}
\end{equation}
with $R_c$ given by
\begin{equation}
\label{rcrit}
R_c \equiv \left( \frac{3 m_p (GM)^{1/2}}{\alpha\kappa k T_i R_i^{\mu}}
\frac{G I \omega_s}{c^2}\right)^{1/(5/2-\mu)}\;.
\end{equation}

The above equation explains why all previous authors were so successful in 
making the disk lie in the neutron star equatorial plane:
the above ratio tends to $0$ at large radii, where the viscosity
is much more effective than LT--forcing. Thus disks are made to lie down 
at large radii. Once the disk material reaches the innermost regions, 
it already lies in the equatorial plane, so that noone will notice that
viscosity may not be effective any longer. 

In particular, this argument explains why we believe the criticisms of 
Lense--Thirring precession in LMXBs (Stella and Vietri 1998) discussed by 
Markovi\'c and Lamb (1998) are not cogent. When seeking {\it global} modes, one 
finds very large damping rates because the {\it global} mode must drag around
ringlets at large radii which are powerfully braked down, because there 
$t_v/t_p \ll 1$, as we now show. 

Markovi\'c and Lamb (1998) eventually decide to consider an isothermal disk
for which $\mu=0$. We then also consider the values they adopt, $J = I 
\omega_s = 3.6\times 10^{48} \;g \; cm^2 \; s^{-1}$, $\kappa=1$, $\alpha=0.1$, 
$T_i = 10^7\; ^\circ K$, $R_i = 9.4\times 10^5\; cm$, and a $1\; M_\odot$ 
accretor. From this we deduce $R_c = 1.0\times 10^8\; cm$. At the same time, 
they take an outer radius given by $x_o = 50$, where $x \equiv \epsilon/\kappa
(R/R_g)^{1/2}$, and $\epsilon=0.2$ is their adopted radiative efficiency.
In physical units, this outer radius is $R_o = 9.3\times 10^9\; cm$. Inserting
the values just deduced for $R_c$  and $R_o$ into Eq. \ref{ratio3} we
find $t_v/t_p \approx 1.2\times 10^{-5}$: as we guessed, their computation 
yields large damping rates because at their large outer radii damping 
by viscosity is much more effective than Lense--Thirring forcing. 
Nor will it do to seek higher order modes, which have more maxima and 
minima: we see in fact from the right--hand--side of Eq. \ref{master} that this 
makes the dissipative term even larger, thus enhancing further still its 
efficiency through the whole radial range.

A more relevant computation, neglecting the clumpy nature of the accretion disk
but still more realistic than the global modes discussed above, would involve 
the time--dependent evolution of a small disk, with an inner free edge and an 
outer edge at small radius being held at fixed misalignment, thusly simulating
the instability that we find in this paper. This will be presented elsewhere.

At small radii we find that $t_v > t_p$. It is convenient to establish
this for a realistic temperature structure. Since we are interested in the 
innermost disk regions, we shall adopt the temperature structure of region A
(Shakura and Sunyaev 1973), 
\begin{equation}
T = 5\times 10^7 \; ^\circ K \alpha^{1/8} m^{1/8} (R/R_g)^{-3/8}\;,
\end{equation}
where $R_g = GM/c^2$. From Eq. \ref{rcrit} we then find
\begin{equation}
\label{rcrittrue}
R_c =  1.1\times 10^8 \; m^{1/17} 
\left(\frac{I}{10^{45}\; g\; cm^2}\right)^{8/17} 
\left(\frac{\nu_s}{300\;Hz}\right)^{8/17} 
\left(\frac{0.1}{\alpha}\right)^{9/17}\; cm \;.
\end{equation}
It is obvious that $R_c$ falls outside region A; we thus have the important 
result, using Eq. \ref{ratio3}, that through the whole region A viscosity 
acts on a slower time--scale than LT--forcing. When this
occurs, then any perturbing agent will be able to excite off--plane
motions, and the viscosity will be unable to damp the motions. 

This bears out our contention (Stella and Vietri 1998) that in both Atoll and 
Z--type sources, if material is excited out of the equatorial 
plane at small radii, then viscosity will be unable to force it 
back down, even for the very large assumed viscosity. Thus {\it any}
excitation acting at small radii, not just the one discussed here, will be
able to induce off-plane motions. This is the reason why work purporting to 
show that LT--precession is unlikely to occur around compact objects is 
irrelevant to our aims: we do not expect that the tilt off the neutron star
equatorial plane is generated at large radii, as was assumed in all
previous works, but rather we showed that it can be excited at 
small radii where it will not be damped.

\section{Discussion and summary}

The processes described above are generic: they apply to every accretion
disk broken up into discrete units, surrounding a magnetized non--aligned 
rotator provided the ordering of time--scales (Eq. \ref{ordering}) 
holds and the blobs have density contrasts $(\rho_b-\rho_d)/ \rho_d \ga 1$. 
The instability is independent of the detailed diamagnetic
properties of the blobs (here enshrined in the parameter $t_d$ which is just 
modulated at the relative frequency $\omega_K-\omega_s$), of their masses 
and densities, their interactions with radiation emitted by the accreting 
source, and the exact form of the viscosity. Provided the hierarchy of 
timescales $t_K \ll t_d \ll t_v$ (Eq. \ref{ordering}) is 
established and blobs have non--negligible density contrasts, it should apply 
to both accreting white dwarfs and neutron stars. 

The existence of resonances is essentially due to the fact that a
spatially periodic magnetic field, such as that due to a single multipole,
will produce a temporally--variable electromagnetic force on a single
blob, as it passes through the various frequency components of the field. 
Minor corrections to the exact locations of the resonances might derive 
from effects we opted not to consider, such as an axially offset field, or
a rotation rate different for different multipoles, as it happens to the
non--dipolar part of the magnetic field of the Earth. Another phenomenon
we neglected to investigate, and which might generate further resonances,
is the non--linear coupling of radial, latitudinal and longitudinal motions. 
It should however be borne in mind that the exact locations of resonances 
might be irrelevant since blobs are expected to have a finite size.

The evolution of the accretion disk after the resonance of Eq. \ref{resonance}
is very difficult to predict; it seems clear enough that blobs will tend to
move closer to the magnetic equatorial plane, and oscillate around it, but
the further evolution is uncertain because the details depend on a very poorly
known quantity, the $z$--axis viscosity. This is the reason why we did not
push the analysis performed in this paper into the non--linear regime. It
seems quite clear that the disk must puff--up, but whether, and when, viscosity
will manage to bring it back to the equatorial plane remains an open question.
Very interesting consequences of the above--discussed 
instability occur in a specific class of accretors, \ie\/ neutron stars 
exhibiting millisecond QPOs. Blobs that have acquired $z$~axis motions will be 
acted upon by the torque which causes Lense--Thirring (1918, LT) precession. We 
have shown here that, in both Atoll and Z--type sources, viscosity is
surely unable to bring blobs back down to the equatorial plane. 
For motions in the plane, the net effect of the instability is to 
induce epicyclic motions, or, given that in $1/r$ potentials all orbits close, 
to perturb the blobs' motions into ellipses. Then periastron precession induced 
by general relativistic effects ought to ensue. 

A related mechanism for the generation of off--plane motions has
been proposed for the Jovian dust ring (Burns \etal\/ 1985). It differs from 
the present one in that the individual units are not blobs of plasma, but
single dust grains with nonzero electric charge, which are then acted upon
by the normal Lorenz force $q \vec{v}^{(rel)}\wedge\vec{B}/c$. It is easy
to see that this differs from our case because the forces differ:
in our case, the acceleration (Eq. 1) is in the plane of the two vectors
$\vec{v}^{(rel)}$ and $\vec{B}$, while in the Jovian case the force is
perpendicular to this plane. This difference has as a consequence that,
in the Jovian case, each magnetic field multipole has its own, single resonance,
while in our case the two multipoles we explored have the same array of 
resonances, and the array is an infinite one.

In short, in this paper we have shown that a simple interaction lifts blobs 
off the equatorial plane of a viscous accretion disk, provided the
accreting object is not an aligned rotator. This occurs in a thin radial
annulus, identified by Eq. \ref{resonance}. This conclusion is rather
general, being independent of the exact form of the drag time--scale, 
of the blobs' diamagnetic properties, and also of the assumed form for the disk 
viscosity. We have also shown why viscosity is unable to damp these
motions (Section \ref{lt}), and that this leads to modulation of the
X--ray flux at the single particle precession frequency, or possibly twice
as much. Thus, the proposed identification of some QPOs in the spectra of 
both Atoll and Z sources as due to LT--precession (Stella and Vietri 1998) 
is made more likely by the existence of a relevant mechanism exciting blobs' 
motions off the equatorial plane.

Helpful comments from F.K. Lamb and J. Imamura are gratefully acknowledged.
This work was begun while one of us (M.V.) was visiting the Institute for
Advanced Study; its Director, John Bahcall, is thanked for the warm hospitality.
The work of L.S. was partially supported through ASI grants.

\vfill

\section*{Figure Captions} 

{\bf Figure 1:} The geometry of the blob--magnetic field interaction. 
The two panels correspond to different phases of the blob orbital
evolution, showing that the acceleration along the $z$ axis (Eq. 1)
alternates in sign.

{\bf Figure 2:} The geometry of the reference systems of Section 2.1. The
$K$ frame is inertial, and its $z$~axis is along the pulsar spin axis. 
The $K'$ frame rotates rigidly with the star; its $z'$~axis is aligned
with the star's symmetry axis of its quadrupole moment. 

\begin{thebibliography}{}

\bibitem{} Bardeen, J.M., Petterson, J.A., 1975, \apj, 195, L65.
\bibitem{} Binney, J.J., Tremaine, S., 1987, {\it Galactic Dynamics},
Princeton Univeristy Press: Princeton, NJ.
\bibitem{} Bodenheimer, P.. 1995, \araa, 33, 199.
\bibitem{} Boyle, C.B., Fabian, A.C., Guilbert, P.W., 1986, Nature, 319, 648.
\bibitem{} Burns, J.A., Schaffer, L.E., Greenberg, R.J., Showalter, M.R., 1985,
Nature, 316, 125.
\bibitem{} Drell, S.D., Foley, H.M., Ruderman, M.A., 1965, J. Geophys.
Res., 70, 3131.
\bibitem{} Field, G.B., 1965, \apj, 142, 531.
\bibitem{} Frank, J., King, A.R., Raine, D.J., 1995, {\it Accretion
Power in Astrophysics}, Cambridge Univ. Press, Cambridge U.K.
\bibitem{} Ghosh, P., Lamb, F.K., 1979, \apj, 234, 296.
\bibitem{} Hatchett, S., Begelman, M.C., Sarazin, C., 1981, \apj, 247, 677.
\bibitem{} Illarionov, A.F., Sunyaev, R.A., 1975, \aap, 216, 578.
\bibitem{} Kallman, T., White, N.E., 1989, \apj, 341, 955.
\bibitem{} King, A.R., 1993, \mnras, 261, 144.
\bibitem{} King, A.R., Lasota, J.P., 1990, \mnras, 247, 214.
\bibitem{} Krolik, J.H., 1998, \apj Letters, in press, astro-ph. n. 9802276.
\bibitem{} Lense, J., Thirring, H., 1918, Physik Z., 19, 156.
\bibitem{} Lightman, A.P., Eardley, D.M., 1974, \apj, 187, L1.
\bibitem{} Markovi\'c, D., Lamb, F.K., 1998, \apj, submitted, astro-ph. n. 
9801075.
\bibitem{} Miller, M.C., 1998, \apj, submitted, astro-ph. n. 9801295.
\bibitem{} Miller, M.C., Lamb, F.K., 1993, \apj, 413, L43. 
\bibitem{} Miller, M.C., Lamb, F.K., 1996, \apj, 470, 1033.
\bibitem{} Miller, M.C., Lamb, F.K., Psaltis, D., 1998, \apj, in press, 
astro-ph. n. 9702090.
\bibitem{} Papaloizou, J.C., Pringle, J.E., 1983, \mnras, 202, 1181.
\bibitem{} Shakura, N.I., Sunyaev, R.A., 1973, \aap, 24, 337.
\bibitem{} Shakura, N.I., Sunyaev, R.A., 1976, \mnras, 175, 613.
\bibitem{} Shapiro, S.L., Teukolsky, S.A., 1983, {\it Black Holes, 
White Dwarfs and Neutron Stars}, J. Wiley and Sons, New York.
\bibitem{} Stella, L. Rosner, B., 1984, \apj, 277, 312.
\bibitem{}Stella, L., Vietri, M., 1998, \apj Lett., 492, L59.
\bibitem{} Stern, D.P., Ness, N.F., 1982, \araa, 20, 139.
\bibitem{} van der Klis, M., 1995, in {\it X--ray Binaries}, eds. W.H.G.
Lewin, J. van Paradijs, E.P.J. van den Heuvel, Cambridge Un. Press, 
Cambridge, p. 252.
\bibitem{} van der Klis, M., 1997, in {\it Proceedings of the Conference on
statistical challenges in modern astronomy II}, Univ. Park PA., USA.
(astro-ph 9704273).
\bibitem{} Vietri, M., Ferrara, A., Miniati, F., 1997, \apj, 483, 262.
\bibitem{} Vishniac, E.T., Diamond, P., 1992, \apj, 398, 561.
\bibitem{} Wynn, G.A., King, A.R., 1995, \mnras, 275, 9.


\end{thebibliography}
\end{document}